\documentclass{appolb1}
\usepackage{epsfig}

\newcommand{\bea}{\begin{eqnarray}}
\newcommand{\eea}{\end{eqnarray}}

\newcommand{\as}{\alpha_s}
\newcommand{\asMZ}{\alpha_s(M^2_Z)}

\begin{document}
\title{A study of QCD coupling constant\\ and power 
corrections  in \\
the fixed target deep inelastic measurements
\thanks{Talk presented at X Int. Workshop on Deep Inelastic Scattering 
(DIS2002), Cracow, 30 April - 4 May 2002. 
A shorter version will appear in the Proceedings.}
}
\author{V.G. Krivokhijine and A.V. Kotikov
\address{Joint Institute for Nuclear
Research, 141980 Dubna, Russia}
}
\maketitle
\begin{abstract}
We reanalyze 
deep inelastic scattering data 
of BCDMS Collaboration by including proper cuts of  ranges
with large systematic errors. 
We perform also the 
fits of high statistic deep inelastic scattering data 
of BCDMS, SLAC, NM and BFP Collaborations  
taking the data separately and in combined way and find good agreement
between these analyses. We
extract the values of both
the QCD coupling constant $\alpha_s(M^2_Z)$ up to
NLO level and of the power corrections to the 
structure function $F_2$. 
\end{abstract}

\section{ Introduction }

The deep inelastic scattering (DIS) leptons on hadrons is the basical
 process to study the values of the parton distribution functions (PDF)
which are universal (after choosing of factorization and renormalization 
schemes) and
can be used in other processes.
The accuracy of the present data for deep inelastic
structure functions (SF) reached the level at which
the $Q^2$-dependence of logarithmic QCD-motivated terms and power-like ones
may be studied separately 
(for a review, see 
\cite{Beneke} and references 
therein).

In the present paper we sketch the results of our analysis \cite{KriKo}
at the next-to-leading order (NLO)
of perturbative QCD for
the most known DIS SF 
$F_2(x,Q^2)$ 
taking into account experimental data \cite{SLAC1}-\cite{BFP} of
SLAC, NM,  BCDMS and BFP Collaborations.
We
stress the power-like effects, so-called twist-4 (i.e.
$\sim 1/Q^2$) 
contributions.
To our purposes we represent the SF $F_2(x,Q^2)$ as the contribution
of the leading twist part $F_2^{pQCD}(x,Q^2)$ 
described by perturbative QCD, 
when the target mass corrections are taken into account,
and the  
nonperturbative part (``dynamical'' twist-four terms):
\vskip -0.5cm
\begin{equation}
F_2(x,Q^2) 
\equiv F_2^{full}(x,Q^2)
=F_2^{pQCD}(x,Q^2)\,
\Bigl(
1+\frac{\tilde h_4(x)}{Q^2}
\Bigr),
\label{1}
\end{equation}
where $\tilde h_4(x)$ is magnitude of twist-four terms.

Contrary to standard fits (see, for example, \cite{Al2000}- \cite{fits}) 
when the direct
numerical calculations based on 
Dokshitzer-Gribov-Lipatov-Altarelli-Parisi 
(DGLAP) 
equation \cite{DGLAP} are used to evaluate structure functions, 
we use the exact solution of DGLAP equation
for the Mellin moments $M_n^{tw2}(Q^2)$ of
SF $F_2^{tw2}(x,Q^2)$
and
the subsequent reproduction of 
$F_2^{full}(x,Q^2)$
at every needed $Q^2$-value with help of the Jacobi 
Polynomial expansion method 
\footnote{We 
note here that there is similar method 
\cite{Ynd}, based on Bernstein polynomials. The method has been used 
in the analyses at the NLO level in \cite{KaKoYaF}
and at the NNLO level in \cite{SaYnd}.}
\cite{Barker,Kri}
(see similar analyses at the NLO level 
\cite{Kri,Vovk}
and at the next-next-to-leading order (NNLO) level and above \cite{PKK}.

In this paper we 
do not present exact formulae of $Q^2$-dependence
of SF $F_2$ which are 
given in \cite{KriKo}. We note only that
the PDF 
at some $Q^2_0$ is theoretical input of our analysis and 
the twist-four term $\tilde h_4(x)$
is considered as a set of free parameters (one constant
$\tilde h_4(x_i)$ per $x_i$-bin):
$\tilde h_4^{free}(x)=\sum_{i=1}^{I} \tilde h_4(x_i)$, 
where $I$ is the number of bins.

\vskip -0.3cm
\section{ 
Fits of $F_2$
}
\label{sec:form}

First of all, we choose the cut $Q^2 \geq 1$ GeV$^2$ in all our studies.
For $Q^2 < 1$ GeV$^2$, the applicability of twist expansion is very
questionable. 
Secondly, we choose  $Q^2_0$ = 90 GeV$^2$ ($Q^2_0$ = 20 GeV$^2$) for the 
nonsinglet (combine nonsinglet and singlet) evolution, i.e.
quite large values of the normalization point
$Q^2_0$: our
perturbative formulae should be applicable at the value of
$Q^2_0$. Moreover, the higher order corrections $\sim \as^k(Q^2_0)$ 
and $\sim (\as(Q^2)-\as(Q^2_0))^k$
($k \geq 2$) should be
less important at these
$Q^2_0$ values.

We use MINUIT program \cite{MINUIT} for
minimization of 
$\chi^2(F_2)$.
We consider
free normalizations of data for different experiments. 
For the reference, we use the most stable deuterium BCDMS data
at the value of energy $E_0=200$ GeV 
($E_0$ is the initial energy lepton beam). 
Using other types of data as reference gives
negligible changes in our results. The usage of fixed normalization
for all data leads to fits with a bit worser $\chi^2$.

\vspace{-0.2cm}
\subsection { BCDMS ~~${}^{12}C + H_2 + D_2$ data }

We start our analysis with the most precise experimental data 
\cite{BCDMS1} obtained  by BCDMS muon
scattering experiment at the high $Q^2$ values.
The full set of data is 762 
points.

It is well known that the original analyses 
given by BCDMS Collaboration itself (see
also Ref. \cite{ViMi}) lead to quite small values 
 $\alpha_s(M^2_Z)=0.113$.
Although in some recent papers (see, for example, 
\cite{Al2000,H1})
more higher values of the coupling constant 
$\alpha_s(M^2_Z)$ have been observed, we think that
an additional reanalysis of BCDMS data should be very useful. 

Based on study \cite{Kri2} 
we proposed in
\cite{KriKo} that 
the reason for small values
of $\alpha_s(M^2_Z)$ coming from BCDMS data was the existence of the subset
of the data having large systematic errors. 
We studied this subject by 
introducing several so-called $Y$-cuts 
\footnote{Hereafter we use the kinematical variable $Y=(E_0-E)/E_0$,
where 
$E$ is
scattering energies of lepton.} 
(see \cite{KriKo}): 
\bea
& &y \geq 0.14 \,\,~~~\mbox{ when }~~~ 0.3 < x \leq 0.4, ~~~~~~~ 
y \geq 0.16 \,~~~\mbox{ when }~~~ 0.4 < x \leq 0.5 \nonumber \\
& &y \geq Y_{cut3} ~~~\mbox{ when }~~~ 0.5 < x \leq 0.6, ~~~~~~
y \geq Y_{cut4} ~~~\mbox{ when }~~~ 0.6 < x \leq 0.7 \nonumber \\
& &y \geq Y_{cut5} ~~~\mbox{ when }~~~ 0.7 < x \leq 0.8 
\label{cut}
\eea
\noindent
and
several $N$ sets  for the cuts at $0.5 < x \leq 0.8$: 
%
\begin{table}[h]
\begin{center}
\begin{tabular}{|c|c|c|c|c|c|c|c|}
\hline
$N$ & 0 & 1 & 2 & 3 & 4 & 5 & 6 \\
\hline \hline
$Y_{cut3}$ & 0 & 0.14 & 0.16 & 0.16 & 0.18 & 0.22 & 0.23 \\  
$Y_{cut4}$ & 0 & 0.16 & 0.18 & 0.20 & 0.20 & 0.23 & 0.24 \\
$Y_{cut5}$ & 0 & 0.20 & 0.20 & 0.22 & 0.22 & 0.24 & 0.25 \\
\hline
\end{tabular}
\caption{The values of $Y_{cut3}$, $Y_{cut4}$ and $Y_{cut5}$.
}\label{tab2}
\end{center}
\end{table}

\begin{figure}[t]
\begin{minipage}[t]{0.48\linewidth}
\begin{center}
\includegraphics[width=3.1in]{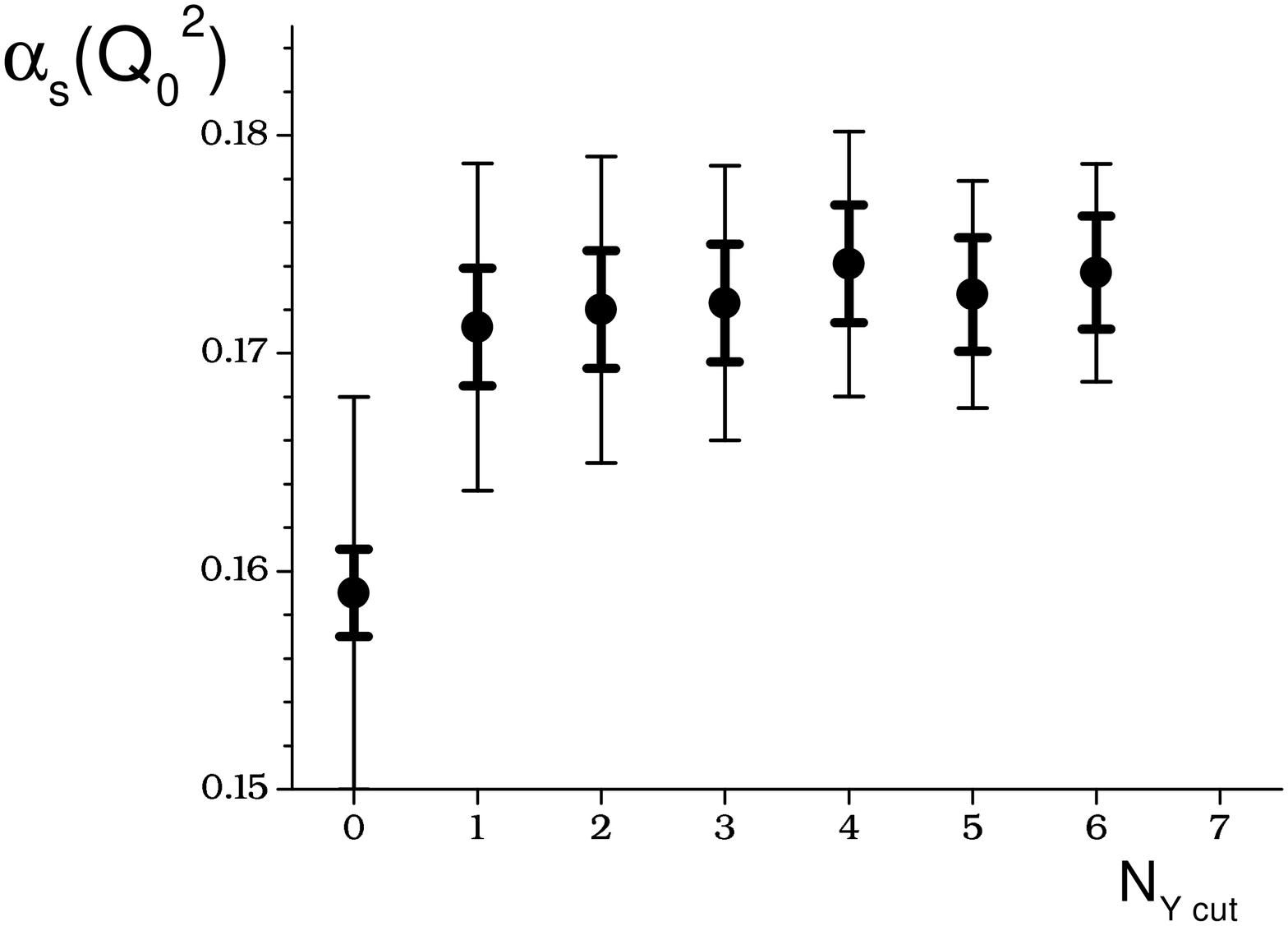} \\
\end{center}
\vskip -0.7cm
\caption{
The study of systematics at different $Y_{cut}$ values
in the fits based on nonsinglet evolution (i.e. when $x \geq 0.25$).
The inner (outer) error-bars show statistical (systematic) errors.
}
 \label{fig:3}
\end{minipage}%
\hspace{0.04\textwidth}%
\begin{minipage}[t]{0.48\linewidth}
\begin{center}
\includegraphics[width=3.1in]{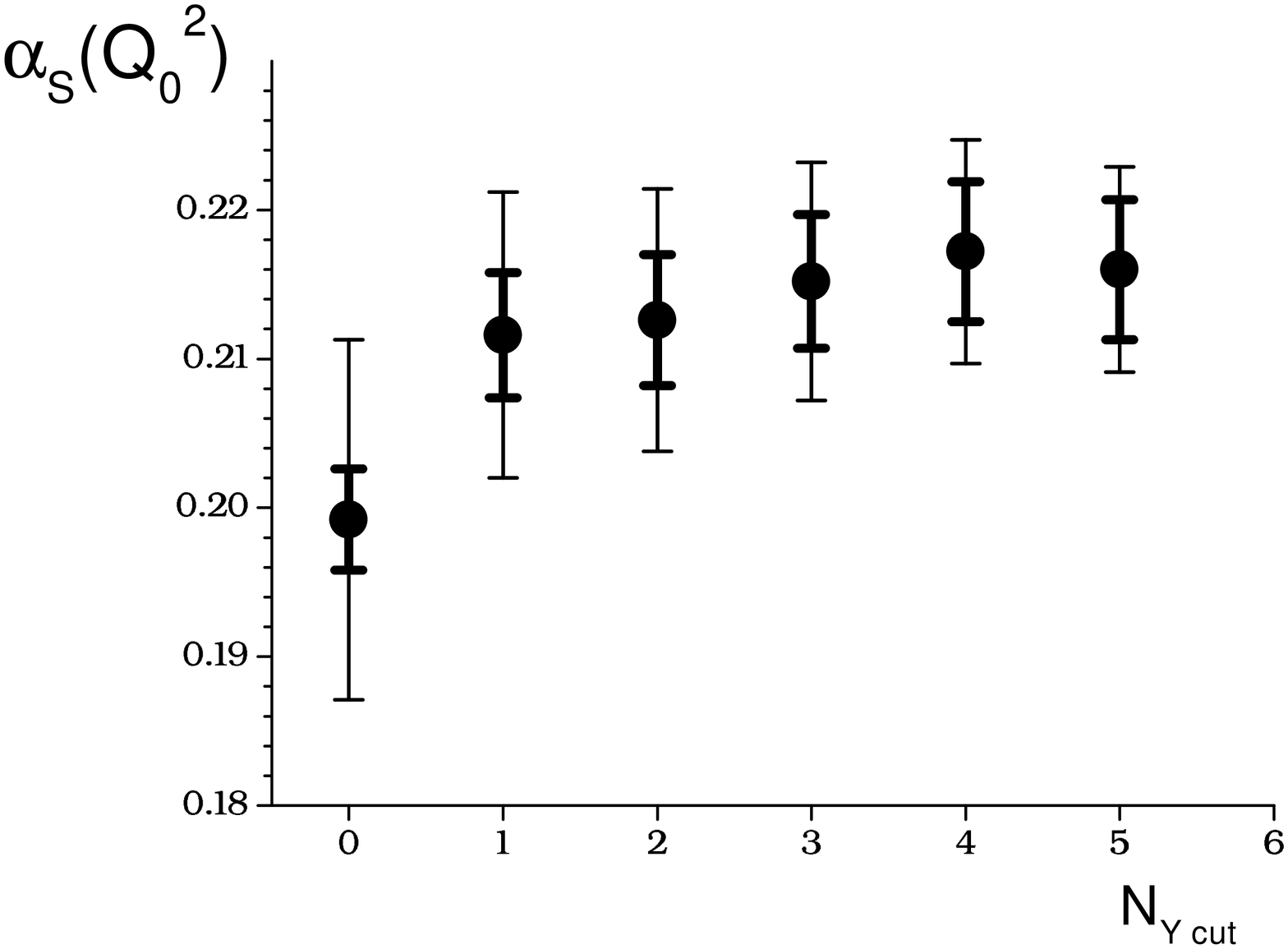} \\
\end{center}
\vskip -0.7cm
 \caption{
All other notes are as in Fig. 1 with two
 exceptions:  the fits based on combine 
evolution
and the points $N_{Ycut}=1,2,3,4,5$ correspond the values
$N=1,2,4,5,6$ in the Table 1.
}\label{fig:4}
\end{minipage}
\end{figure}

\vskip -0.5cm
From 
the Figs. 1 and 2 we can see that the $\alpha_s$ values are obtained
for $N=1 \div 6$ of $Y_{cut3}$, $Y_{cut4}$ and $Y_{cut5}$ are very stable and
statistically consistent. 

\vskip -0.5cm
\subsection{ SLAC, BCDMS, NM and BFP data }
\label{subsec:revi1}
After these $Y$-cuts have been incorporated (with $N=6$) for BCDMS data, 
the full set of combine data is 1309 
points.
The results of the fits 
are compiled in Summary.

\begin{figure}[t]
%
\begin{center}
\psfig{figure=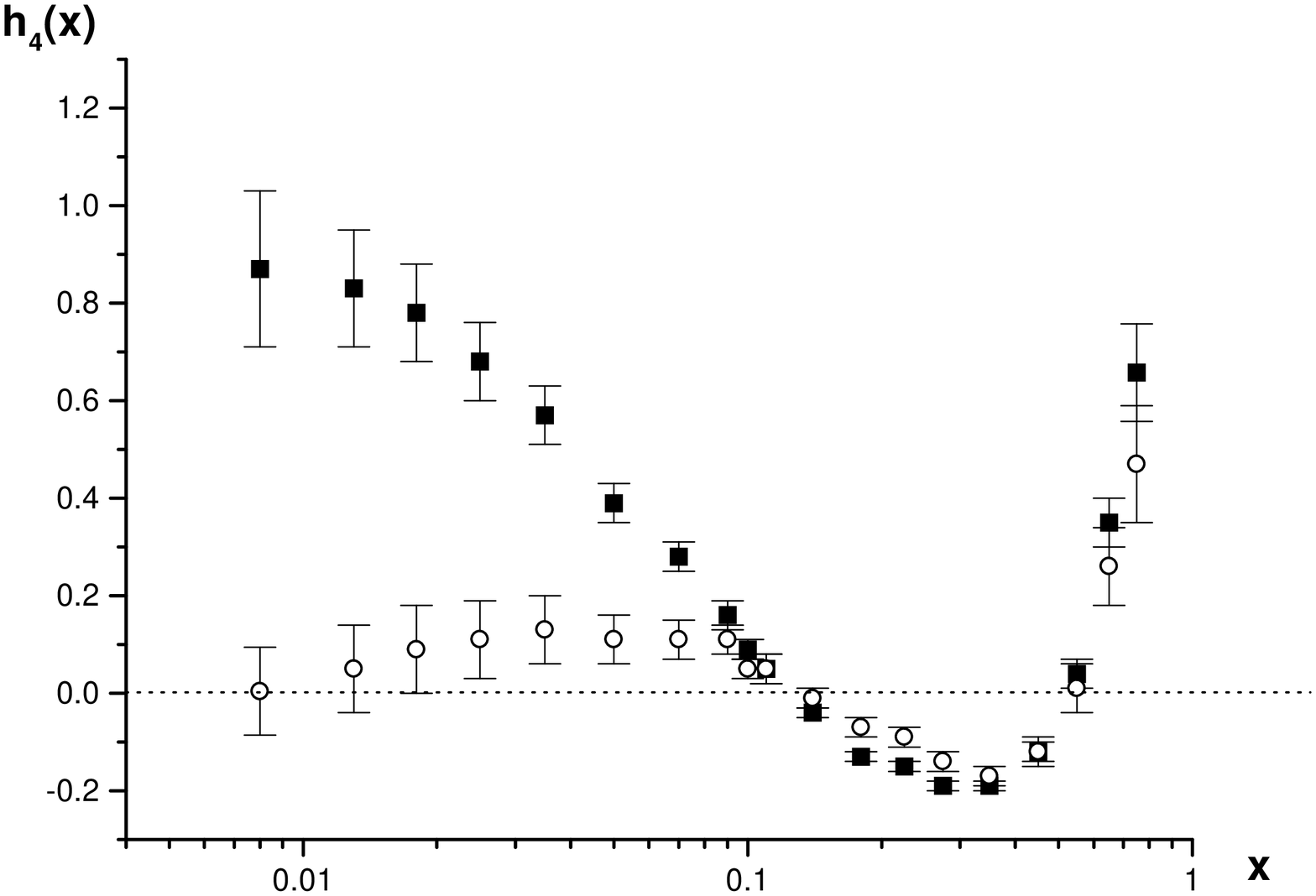,width=16cm,height=13cm}
\end{center}
\vskip -1cm
\caption{
The values of the twist-four terms. The black and white points correspond 
to the small-$x$ asymptotics $\sim x^{-\omega}$ of sea quark and gluon
distributions
with $\omega =0$ and
$\omega =0.18$, respectively. The statistical errors are displayed only. 
}
\label{HT}
\end{figure}

\section{ Summary }

We
have demonstrated several steps of our study \cite{KriKo}
of the $Q^2$-evolution of DIS structure function $F_2$ fitting all
fixed target experimental data.

From the fits we have obtained the value of the normalization 
$\asMZ$
of QCD coupling constant. First of all, we have reanalyzed the BCDMS data 
cutting the range with large systematic errors. As it is possible to see
in 
the Figs. 1 and 2, 
the value of $\asMZ$ rises strongly when
the cuts of systematics were incorporated. In another side, 
the value of $\asMZ$ does not dependent on the concrete type of the
cut within 
modern statistical errors.

Fitting SLAC, BCDMS, NM and BFP data,
we have found in \cite{KriKo} that at $Q^2 \geq 10 \div 15$ GeV$^2$ 
the formulae of pure perturbative
QCD (i.e. twist-two approximation together with target mass corrections)
are in good agreement with all data
\footnote{We note that at small $x$ values, the perturbative QCD
works well starting with $Q^2 = 1.5 \div 2$ GeV$^2$
and higher twist corrections are important only at very low $Q^2$:
$Q^2 \sim 0.5$ GeV$^2$ (see \cite{Q2evo,HT,Slope} and references therein).
As it is was observed in \cite{DoShi,bfklp} (see also discussions in
\cite{Q2evo,HT,BoAnd}) the good agreement between perturbative QCD and
experiment seems connect with large effective argument of coupling
constant at low $x$ range.}. 
When we have added twist-four corrections, we have very good agreement
between QCD (i.e. first two coefficients of Wilson expansion)
and the data starting already with $Q^2 = 1$ GeV$^2$, where the Wilson
expansion should begin
to be applicable.
 The 
results for  $\asMZ$ are very similar (see \cite{KriKo}) for the 
both types of analyses and have the following form:
\bea
\as(M_Z^2) &=& 0.1177 \pm 0.0007 ~\mbox{(stat)}
\pm 0.0021 ~\mbox{(syst)} \pm 0.0009 ~\mbox{(norm)}, 
\label{re2s}
\eea
where the symbols ``stat'', ``syst'' and 
``norm'' mark the 
statistical error, systematic one and the 
error of normalization of experimental data.

We would like to note that we have good agreement also with the analysis 
\cite{H1} of
combined H1 and BCDMS data, which has been given by H1 Collaboration very 
recently. 
Our results for $\as(M_Z^2)$ are in good agreement also with 
the average value for coupling constant,
presented in the recent studies (see \cite{Al2000,LEP}
and references therein) and in
famous 
reviews \cite{Breview}.

At the end of our paper we would like to discuss the contributions of 
higher twist corrections. 
In our study \cite{KriKo}
we have reproduced well-known $x$-shape of
the twist-four corrections at the large and intermediate values of 
Bjorken variable $x$ (see the Fig. 3 and \cite{ViMi,KriKo}).

Note that there is a small-$x$ rise of the magnitude of
twist-four corrections, when we use flat parton distributions at $x \to 0$.
As we have discussed in Ref \cite{KriKo},
there is a strong correlation 
\footnote{ 
This correlation comes because of very limited numbers of 
experimental data used here lie at the low $x$ region. Indeed, only the NMC
experimental data contribute there. 
We hope to incorporate the HERA data \cite{H1,ZEUS}
in our future investigations.}
between the small-$x$ behavior of twist-four
corrections and the type of the corresponding asymptotics of the 
leading-twist parton distributions. The possibility to have  a
singular type of
the asymptotics leads (in our fits)
to the appearance of the rise of sea quark and gluon 
distributions as $\sim x^{-0.18}$ at low $x$ values,
that is in full agreement with low $x$ HERA data
and with theoretical studies \cite{Q2evo,bfklp}.
 At this case
the rise of the magnitude
of twist-four corrections is completely
canceled. This cancellation
is in full agreement
with theoretical and phenomenological studies 
(see \cite{Q2evo,Bartels1,HT}).

\vskip 0.2cm
{\bf Acknowledgments.}
The study is supported in part by 
the Heisenberg-Landau program.
A.V.K.  would like to express his sincerely thanks to the Organizing
  Committee of the X Int. Workshop on Deep Inelastic Scattering 
(DIS2002)  for the kind 
invitation
and  for fruitful discussions.
He was supported in part by Alexander von Humboldt
fellowship and INTAS  grant N366.

\end{document}